# An Integrated Framework for Performance Analysis and Tuning in Grid Environment

Ajanta De Sarkar, Sarbani Roy, Sudipto Biswas and Nandini Mukherjee
Department of Computer Science and Engineering
Jadavpur University
Kolkata - 700 031
Email: nmukherjee@cse.jdvu.ac.in

*Abstract:* In a heterogeneous, dynamic environment, like Grid, post-mortem analysis is of no use and data needs to be collected and analysed in real time. Novel techniques are also required for dynamically tuning the application's performance and resource brokering in order to maintain the desired QoS. The objective of this paper is to propose an integrated framework for performance analysis and tuning of the application, and rescheduling the application, if necessary, to some other resources in order to adapt to the changing resource usage scenario in a dynamic environment.

## 1. Introduction

Traditionally, performance monitoring and tuning is a cyclic and feedback guided process – every time it is needed to analyse and tune the performance of the application and to send back the application for execution. However, unlike traditional architectures, post-mortem analysis is of no use in dynamic distributed environments, like Grid. For successful performance analysis in such environments, real time performance data needs to be captured. Resource brokering and allocation on the basis of these real-time performance data is a major issue in Grid environment. Dynamic policy selection in response to current resource availability and application demands is also one of the important requirements.

In this paper, we present an integrated framework for analysis and tuning the application's execution performance and dynamically rescheduling the application in order to adapt with the changing execution environment. The framework is a part of a multi-agent system [14] which supports performance-based resource allocation for multiple concurrent jobs executing in a Grid environment.

## 2. Related Work

Several tools for measuring or analyzing performance of serial/parallel programs in distributed environment have been developed so far; these include SCALEA [16], Pablo [12], EXPERT [18], etc. There are many other existing tools for performance monitoring and analysis in grid environments. In the ASKALON project, a set of four tools are involved in performance prediction, analysis, and measurement and data collection. SCALEA-G [17] is another tool that is based on the concept of Grid Monitoring Architecture (GMA) [15]. It provides an infrastructure for OGSA [4] and supports performance analysis of a variety of Grid services including computational resources, networks, and applications. GRM and R-GMA [10] are used for on-line monitoring of parallel applications running on the grid. These tools collect trace information particularly for message passing parallel applications. Pulse [11] is developed within the EU DataGrid project as an analysis and presentation tool. Pulse is a framework to compose an analyzing chain for grid monitoring data about services or resources. In the Peridot [5] project, distributed performance analysis system is composed of a set of analysis agents. The agents at the lowest level in the hierarchy are responsible for the collection and analysis of performance data from one or more nodes upon request from monitors.

While our work also proposes the use of a hierarchically organized set of analysis agents, a strict categorisation of these analysis agents is proposed and the functions of each of these agent categories are clearly defined. Moreover, our main focus is on the implementation of an integrated framework in which different categories of agents will collaborate not only for performance analysis, but also for using the analysis results and thereby improving the performance of the application in real-time. Thus, we integrate local tuning agents on each of the grid resources for dynamically changing the application codes. Our framework also supports performance-driven job migration in a grid environment.

## 3. An integrated Agent Framework

The environment which we have selected for implementation of our framework is depicted in Figure 1. In this environment [8] Grid resources can be clusters or SMPs or even workstations of dissimilar configuration, but all of these are tied together through a grid middleware layer. A Grid site comprises a collection of all these local resources, which are geographically located in a single site. All these Grid sites, which are mutually agreed to share resources located in several sites form a Global Grid. Global grid is responsible for multiple grid resource registries and grid security services through mutually distrustful administrative domains.

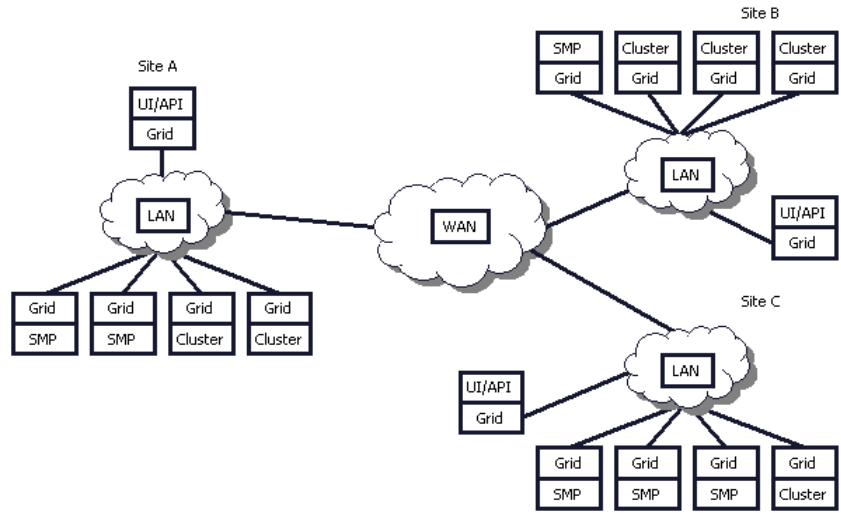

**Figure 1 Grid Environment**

In the above environment, our framework undertakes two different actions for improving the performance of a running application. The first action is tuning a part of the application locally, whereas the second action is rescheduling the application to a different resource in case the current resource fails to keep its promise. The actions are taken either by a *Tuning Agent* (local tuning) or by a *Job Controller* and a *Job Execution Manager Agent* (rescheduling) [14] of the agent framework. However, any of these actions are taken on the basis of some performance data collected by the *Analysis Agents* organized in a hierarchy. These analysis agents are divided into the following four logical levels of deployment in descending order: (1) Grid-level Agent (GA), (2) Grid Site-level Agent (GSA), (3) Resource-level Agent (RA) and (4) Node-level Agent (NA). In the following paragraphs, we describe in detail the activities of each type of analysis agents and also their interaction with other agents.

**Node-level Agents (NA) -** *Node-level Agents* are located at each node, i.e. on a workstation, on an SMP or on each node of a cluster. An NA interacts with a performance monitor at the lowest level and collects performance data in order to analyse the performance of the node or the execution performance of a job running on that node. Thus it can provide information like CPU usage, memory access pattern or even load imbalance in case of an SMP. In case the execution performance of a job degrades due to a performance problem that can be solved locally, the NA raises a warning message. Immediately a tuning agent is invoked (discussed later) for local tuning of the job. For example a resource-crunch job may be allowed to use more resources at the run-time or the load-imbalance effect may be reduced by dynamically changing the local scheduling strategy.

**Resource-level Agents (RA) -** These agents seat at the resource level. For example, a cluster requires one RA that interacts with all the NAs sitting on the cluster nodes. It gathers performance data from these NAs and analyses the overall performance of the cluster. An RA can detect problems like load imbalance in the cluster. RAs also interact with the *JobExecutionManager Agents(JEM)* and keep them informed about the current status of the resource and execution performance of each job running on the resource. Depending on this information only, the *JobExecutionManager* and the *JobController* Agents may take decisions about the migration of any job to any other resources [14]. Individual SMPs and workstations participating in a Grid also admit an RA in their environment. These RAs do not require to carry out any performance analysis in addition to what is done by the NAs; however they act as intermediaries between the NAs and the JEMs and assist the JEMs to take appropriate decisions.

**Grid Site-level Agents (GSA)** - A GSA keeps track of all the resources in a particular Grid-site. It interacts with the RAs and gathers summary reports regarding execution of various jobs on different resource providers at the particular site. A GSA is useful to identify any fault in any of the resources and also to check whether any of the resources is overloaded. A JobController Agent interacts with the GSAs (at the local site or at a remote site) at the time of migration of a job, in order to find the next suitable resource. The GSA checks the current loads of all resources that can meet up the requirement of a job, and sends this information to the JobController Agent. On the basis of this information, the JobController Agent takes decision regarding where to migrate a job. In addition to this, a GSA can proactively inform the JobController Agent regarding any fault or any performance problem at a particular resource and, thereby, allowing the agent to determine its next course of action.

**Grid-level Agents (GA)** - A GA does not take part in the execution performance analysis of jobs; rather it focuses on the overall health of the Grid. Thus, it analyses information like, whether every resource-owner has been successful to keep their promises and whether every job has received the computational services as they have been assured for. This analysis is based on the data collected from the GSAs and the JobController Agents. Thus each GSA and each JobController Agent must regularly update the GA for this analysis.

Figure 2 demonstrates interaction among all the agents participating in the performance analysis and tuning process.

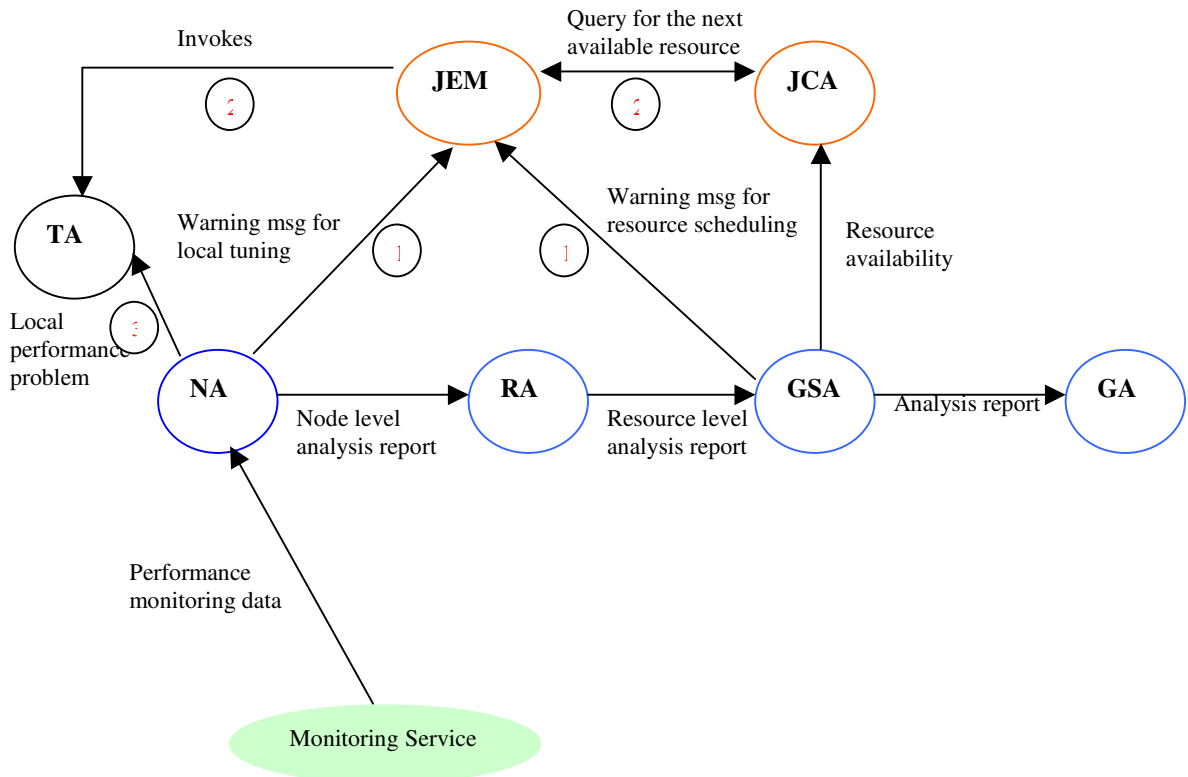

**Figure 2 Agent Interaction Diagram**

## 4. Tuning Scenarios

In this paper two scenarios are presented to demonstrate the use of the proposed framework. Also we present the preliminary results of our implementation in this section.

**Scenario 1** – A parallel code (with OpenMP directives) starts executing using a single thread on a multi-processor machine and a Node-level Analysis Agent monitors its performance by profiling the code. The application code is suitably instrumented to capture a set of events, data regarding the occurrences of events are stored in a data buffer and NAs use a *pull model* for obtaining data from the data buffer. This data is then analysed by the NAs in order to discover the occurrence of any performance problem. At this point the NA may identify that performance needs to be improved and discovers that performance can be improved by creating more threads and actually executing part of the code parallely. Thus the local Tuning agent is invoked and the number of threads is increased at run-time. Figure 3(a) shows the performance improvement for a test code (for adding two matrices) when number of threads was increased from one to two. Further improvement could be observed if we had used more threads for parallel execution of the code.

**Scenario 2** – A parallel code is executing on a server. The Node-level Analysis Agent discovers that desired performance cannot be obtained because of resource limitation on the particular node. A GSA which regularly interacts with all the NAs / RAs at a particular site obtains this information from the NA. An NA or an RA can also proactively send a message to the GSA indicating a fault or overloading situation (thus, we use a pull as well as a push model of data collection). As soon as a GSA recognizes a situation where job migration is necessary, it informs the concerned JEM Agents by sending appropriate warning messages. The JEM Agent, in order to find a suitable resource, consults the JobController Agent. The JobController Agent first checks the resource availability, runs a resource selection algorithm (not within the scope of this paper) and advises the JEM Agent regarding its next course of action. The JobController Agent queries the GSAs regarding the current status of a set of resources. As the resources may be located at different sites, the JobController Agent may need to check with different GSAs in the Grid. GSAs, depending upon the information gathered from the NAs / RAs, dynamically decide the status of each resource in the set and return this information to the JobController Agent.

Figure 3(b) demonstrates the effect of migration of a simple sorting program (initially scheduled on a two-processor server, referred to as Server1) to a 16 processor server, referred to as Server2. Although in our results we use only four processors on Server2. The performance of the migrated code is compared with the performance of an equivalent serial code and the performance of a parallel code that continued to execute on Server1. We demonstrate the effect with different data sizes and as we see, every time the performance improves quite noticeably. More detailed implementation of this part has been discussed in [14].

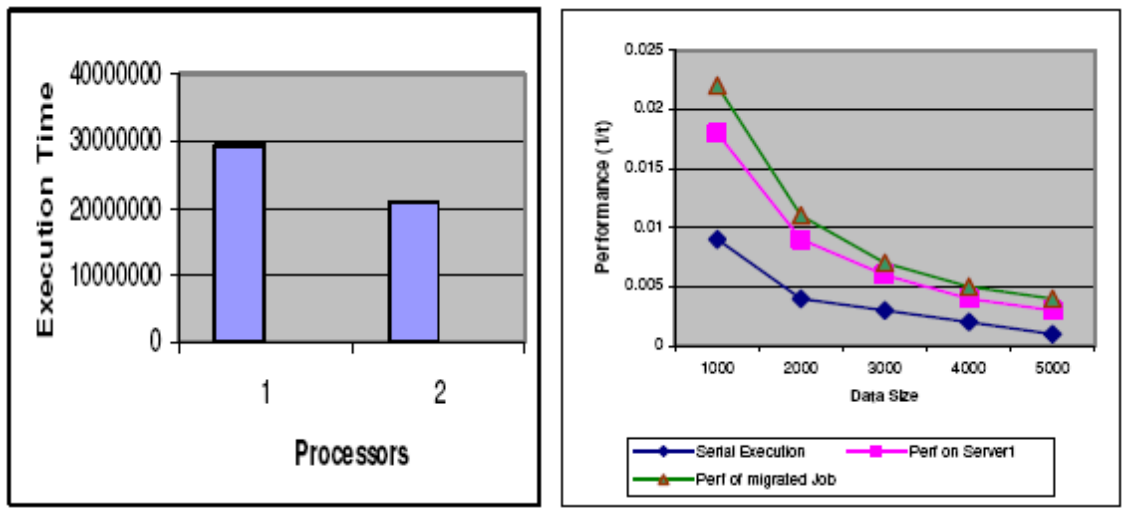

Figure 3 (a) Performance Improvement for Matrix Addition Code in Scenario 1 (time is given in seconds), (b) Performance Improvement for Sorting Code in Scenario 2

**Accounting and Auditing**

Any performance problem noted by the NAs, RAs and GSAs is communicated to the Grid Agent (GA). The lower level agents proactively register this information with the GA. The GA may later use this information for accounting and auditing purpose. However, accounting and auditing are different issues and our work does not focus on these issues. Therefore, we restrict our discussion in this regard.

## 5. Conclusion

In this paper, we propose an integrated framework for performance analysis of applications executing on large distributed systems and also dynamically improving their execution performances. Interaction and exchange of information among the agents for collecting data, analyzing and improving the performance have also been discussed. Two usage scenarios of the framework and preliminary results of partial implementation of the framework have been described.